\documentstyle[12pt]{article}
\newcommand{\bce}{\begin{center}}
\newcommand{\ece}{\end{center}}
\newcommand{\beq}{\begin{equation}}
\newcommand{\eeq}{\end{equation}}
\newcommand{\bea}{\vspace{0.25cm}\begin{eqnarray}}
\newcommand{\eea}{\end{eqnarray}}

\newcommand{\ba}{\begin{array}}
\newcommand{\ea}{\end{array}}

\newcommand{\doublespace}{
    \renewcommand{\baselinestretch}{1.6}\large\normalsize}

\def\lsim{\mathrel{\rlap{\lower4pt\hbox{\hskip1pt$\sim$}}
    \raise1pt\hbox{$<$}}}         
\def\gsim{\mathrel{\rlap{\lower4pt\hbox{\hskip1pt$\sim$}}
    \raise1pt\hbox{$>$}}}         

\def\lsim{\mathrel{\rlap{\lower4pt\hbox{\hskip1pt$\sim$}}
    \raise1pt\hbox{$<$}}}         
\def\gsim{\mathrel{\rlap{\lower4pt\hbox{\hskip1pt$\sim$}}
    \raise1pt\hbox{$>$}}}         

\def\PRL{{Phys. Rev. Lett.} }
\def\PRD{{Phys. Rev.} D }

\def\ZPC{{Z. Phys.} C }
\def\ZPA{{Z. Phys.} A }
\def\PTP{{Progr. Th. Phys. }}
\def\LNC{{Lett. al Nuovo Cimento} }

  \advance\oddsidemargin  by -1in
  \advance\evensidemargin by -1in
  \advance\topmargin      by -0.5in
\def\lsim{\mathrel{\rlap{\lower4pt\hbox{\hskip1pt$\sim$}}
    \raise1pt\hbox{$<$}}}         
\def\gsim{\mathrel{\rlap{\lower4pt\hbox{\hskip1pt$\sim$}}
    \raise1pt\hbox{$>$}}}         

\def\beq{\begin{equation}}
\def\endeq{\end{equation}}
\def\arr{\begin{eqnarray}}
\def\endarr{\end{eqnarray}}
\makeindex

\doublespace
\begin{document}


\begin{center}
{\Large \bf Baryons Electromagnetic Mass Splittings  in

Potential Models}

{\large \em  {\underline {Marco Genovese}$^a$ \footnote{ \small                     
Supported by the EU Program ERBFMBICT 950427} }, 
Jean-Marc Richard$^a$, Bernard Silvestre-Brac$^a$ and
 K{\'a}lm{\'a}n Varga$^b$}

{\small $^a$Institut des Sciences Nucl\'eaires, 
Universit\'e Joseph Fourier--CNRS-IN2P3\\
53, avenue des Martyrs, F--38026  Grenoble \\
$^b$ I.N.R. of the Hungarian Academy of Sciences, 
Debrecen, PO Box 51, Hungary\\
and RIKEN, Hirosawa 2-1, Wako, Saitama 35101, Japan}

\vspace{0.4cm}
{\bf Abstract}\\
\end{center}
\noindent {\small We study electromagnetic mass splittings of charmed 
baryons.  
We point out  discrepancies among theoretical predictions 
in non-relativistic potential models;  none of these predictions
seems supported by experimental data.
A new calculation is presented.}

 \bigskip

 \bigskip

Quite a successful phenomenology has been obtained
using non-relativistic 
potential models built in order to describe the low-energy limit 
of QCD. 
Among the various observables studied in such models, a large
amount of work was devoted to the electromagnetic mass
differences \cite{MiY}.
In general a good agreement was obtained for nucleon, $\Sigma$ and $\Xi$ baryons 
 and  predictions for charmed baryons were
supplied (see table 1).
Albeit the predictions for charmed baryons were quite 
dispersed, some general feature was anyway common to most of 
the models, as
$\Sigma^+_c-\Sigma_c^0 \lsim \Sigma^{++}_c-\Sigma_c^0$.

Surprisingly enough, when data about charmed baryons finally 
appear, none of these models  came out in agreement with them.
Because of the poor amount of experimental data and large errors, we 
cannot really exclude that future experimental determination 
could change the situation. However, aimed by 
this failure, we have decided to investigate 
isospin-violating mass differences of baryons  in a successful potential 
model, including all possible contributions.
We wish to understand whether the failure of potential models arises
from technical problems (as having neglected some contributions,
or having used  perturbative procedures unproperly, etc.) 
or reveals some intrinsic limitation of this approach.

In the following we will use the potential model AL1 \cite{SBS} 
supplemented with electric and dipole--dipole magnetic interactions. 
The difference of masses between $u$ and $d$ quarks has been fixed 
to reproduce the $n-p$ and $\Sigma^{-} - \Sigma^{+}$ 
mass splittings. 

Our  result is that, while $\Sigma^{-} - \Sigma^{0}$,
$\Xi^{-} - \Xi^{0}$ and $\Delta^{0} - \Delta^{++}$ and even 
(within large errors) the splitting of excited states $\Sigma(1385)$
and $\Xi(1530)$ come out in good 
agreement with the experimental data \cite{PDB}, some problems appear
for charmed baryons (see table 2).
In fact, while the experimental datum $\Sigma_c^{++} - \Sigma_c^{0}=
0.8 \pm 0.4$ MeV is well reproduced, one finds a  
negative $\Sigma_c^{+} - \Sigma_c^{0}$, at 
variance with the experimental datum $ 1.4 
\pm 0.6$ MeV. The result $\Xi_c^{+} - \Xi_c^{0} = 2.2$ MeV
is smaller than the PDG average
$6.3 \pm 2.3 $ \cite{PDB}, but agrees rather well with a new 
determination at CLEO, $ 2.5 \pm 1.7 \pm 
1.1$ \cite{CLEO}.
Reasonable changes of light quark masses do not modify substantially 
this situation.

The problem with the $\Sigma_c$ multiplet 
raises the question whether some contribution has been 
neglected. 
For example some models, though reproducing quite well the excitation
energies, fail in providing good absolute mass predictions unless an
empirical three-body interaction is introduced. Its form is rather    
arbitrary and is chosen only for the sake of 
simplicity as \cite{SBS}
$
D_3 +  A_3 (m_1 m_2 m_3)^{-b_3}
$.
As it depends on masses, this term gives a contribution
to electromagnetic mass splittings as well. However, how is evident by 
inspecting this term, the contribution to $\Sigma_c^{+} - 
\Sigma_c^{0}$ goes in the wrong direction for solving the splitting
problem.
In fact, our numerical study shows that one obtains reasonable electromagnetic mass splittings
for light baryons, but  the situation 
for charmed baryons slightly deproves with 
$\Sigma_c^{+} - \Sigma_c^{0} \simeq -0.7$ MeV, and 
$\Xi_c^{+}-\Xi_c^{0} \simeq 2$ MeV.
One could think of more complicated three-body interactions,
but their form remain completely arbitrary and somehow the need
of introducing complicate multi-body interactions would cast doubts
on the applicability of potential models.

One can consider also  the running of 
$\alpha_s$, which is smaller when heavy quarks appear,
for the scale is proportional to the masses involved. 
Such an effect would decrease the coupling of the spin--spin 
term and does not go in the right direction for changing the 
order within the $\Sigma_c$ multiplet.
 
We also ask ourselves whether instantonic interactions 
could improve the situation. The non-relativistic form of this 
contribution has be evaluated in 
\cite{Dorokhov} (the value of the coupling must be fixed
phenomenologically). 
However the effect of
this interaction is inversely proportional to the quark masses 
and vanishes for a quark pair with spin $1$, thus will not contribute 
substantially to $\Sigma_{c}$ mass splittings; it gives, 
however, a positive contribution to $\Xi_c^{+} - \Xi_c^{0}$.

In conclusion\footnote{We will not discuss potentials related to meson 
exchange, for which an extension to charmed baryons is problematic.}
 we find that, albeit a good agreement with
light quark baryons, it is  practically impossible to explain the 
splitting pattern of  charmed baryons in potential models based on
one--gluon exchange.
More precise data are however required before drawing firm conclusions about 
the relevance of such models to describe the confining region of QCD.

\pagebreak

-{Table 1}. {\small Predictions of different  models for charmed baryons 
electromagnetic mass splittings}
$$\vbox {\halign {\hfil #\hfil &&\quad \hfil #\hfil \cr
\cr \noalign{\hrule}
\cr \noalign{\hrule}
\cr

&Model \cite{MiY}  && $\Sigma_c^{++} - \Sigma_c^{0}$&& $\Sigma_c^{+} - 
\Sigma_c^{0}$ && $\Xi_c^{0} - \Xi_c^{+}$&\cr
\cr \noalign{\hrule}
\cr

& Itoh  && 6.5  && 2.4 && 2.51 &\cr

& Ono  && 6.1 && 2.24 && 1.77&\cr

& Lane and Weinberg  && --6 && --4 && 4 &\cr

& Chan  && 0.4 && --0.7 && 3.2 &\cr

& Lichtenberg  && 3.4 && 0.8 && 1.1 &\cr

& Kalman and Jakimow  && --2.7 && --2.24 && 3.6 &\cr

&Isgur  && --2 && --1.8 && &\cr

& Richard and Taxil I  && 3 && 1 && 0&\cr

&  \phantom{Richard and Taxil}  II  && --2 && --2 && 2&\cr
  
\cr \noalign{\hrule}
}}$$

-{Table 2}. {\small Our predictions of 
electromagnetic mass splittings (in MeV, upper row)
 with $m_u=327$MeV and $m_d=338$ 
MeV compared with experimental data (lower row).\footnote{The $u-d$
mass difference is somehow larger than the common wisdom current quarks 
one, however the constituent quarks $u-d$ mass difference can be 
in principle different from the former.}
$$\vbox {\halign {\hfil #\hfil &&\quad \hfil #\hfil \cr
\cr \noalign{\hrule}
\cr \noalign{\hrule}
\cr
$ n - p$ && $\Sigma^{-} - \Sigma^{0}$&& $\Sigma^{-} - 
\Sigma^{+}$ && $\Xi^{-} - \Xi^{0}$&&$\Delta^0-\Delta^{++}$&&
$\Delta^+-\Delta^{++}$&\cr

 1.24 && 5.24 && 8.67 && 7.46 && 2.54 && 0.36&\cr

1.293318 $\pm$ 0.000009 && 4.88 $\pm $ 0.08 && 8.09 $\pm$ 0.16 && 
6.4 $\pm$ 0.6 && 2.7 $\pm$ 0.3 && &\cr
\cr \noalign{\hrule}
\cr

$\Delta^- -\Delta^{++}$&&${\Sigma^{*} } ^0 - {\Sigma ^{*} } ^+$&&$
{\Sigma ^{*} }^-  - {\Sigma ^{*} }^0 $&&${\Xi ^{*} } ^{-}   - 
{\Xi ^{*} }^0$&& &\cr

6.55 && 1.9 && 3.8 && 3.6 && && &\cr

 && --4 to 4 && 2.0 $\pm$ 2.4 && 3.2 $\pm$ 0.6 && && &\cr
\cr \noalign{\hrule}
\cr

 $\Sigma_c^{++} - \Sigma_c^{0}$&& $\Sigma_c^{+} - 
\Sigma_c^{0}$ && $\Xi_c^{0} - \Xi_c^{+}$
&& $\Sigma_b^{+} - \Sigma_b^{-}$&& $\Sigma_b^{0} - 
\Sigma_b^{-}$ && $\Xi_b^{-} - \Xi_b^{0}$&\cr

 1.20 && --0.36 && 2.83 && --3.58 && --1.94 && --5.39&\cr

 0.8 $\pm 0.4$ && 1.4 $\pm$ 0.6 && 6.3 $\pm$ 2.3 && && && &\cr

\cr \noalign{\hrule}
}}$$

\end{document}